# A planar Runge-Lenz vector

S. G. Kamath[*]

Department of Mathematics, Indian Institute of Technology, Madras 600 036, India

**Abstract** :Following Dahl's method an exact Runge-Lenz vector **M** with two components $M_1$ and $M_2$ is obtained as a constant of motion for a two particle system with charges $e_1$ and $e_2$ whose electromagnetic interaction is based on Chern-Simons electrodynamics. The Poisson bracket $\{M_1,M_2\} \neq L_z$ but is modified by the appearance of the product $e_1 e_2$ as central charges.

*e-mail: kamath@acer. iitm. ernet. in

A characteristic feature of the non-relativistic Kepler problem is that there exists apart from the conservation of energy(E) and the angular momentum vector (**L**) another conserved quantity, namely the Runge-Lenz vector (**A**). A textbook[1] derivation of **A** begins with the equation of motion of a mass m under the central force $\bm{F} = f(r)\dfrac{\bm{r}}{r}$, namely, $\dfrac{d\bm{p}}{dt} = f(r)\dfrac{\bm{r}}{r}$, leading to

$$\frac{d}{dt}(\bm{p} \times \bm{L}) = -mf(r)r^2 \frac{d}{dt}\left(\frac{\bm{r}}{r}\right) \tag{1}$$

as $\dfrac{d\bm{L}}{dt} = 0$. With $r^2 f(r) =$ a constant(say - k), implying thereby an inverse square law of force, eq. (1) immediately yields $\bm{A} = \bm{p} \times \bm{L} - mk\dfrac{\bm{r}}{r}$; this is the Runge-Lenz vector. With the definition $\bm{K} = (-2mE)^{1/2} \bm{A}$, one easily obtains the Poisson bracket relations

$$\{\bm{K}, E\} = 0, \{\bm{L}, E\} = 0, \{L_i, L_j\} = \varepsilon_{ijk} L_k, \{L_i, K_j\} = \varepsilon_{ijk} K_k, \{K_i, K_j\} = \varepsilon_{ijk} L_k \tag{2}$$

Thus there exists an internal symmetry associated with the non-relativistic Kepler problem with the invariance group being isomorphic to the 4-dimensional rotation group $O_4$. Until recently, the presence of this internal symmetry had not been tied to a generally accepted invariance principle. In other words, the phenomenological derivation implied in eq. (1) begs the question of whether the Runge-Lenz vector has a deeper physical origin. Specifically, the question is if there is a space-time transformation, the invariance of the Lagrangian for the Kepler problem under which directly leads to the conservation of **A**.

An affirmative answer to this question was recently obtained by Dahl[2] by regarding the Kepler problem as the zero-order description of a relativistic two-body problem; or, as emphasized by Dahl[2], it is absolutely necessary to investigate the relativistic two-body problem in order to discover the connection between the dynamical symmetry of the non-relativistic Kepler problem and special relativity.

Of special relevance to this paper is the fact [3] that for the Kepler problem the angular momentum vector $\mathbf{L} = \mathbf{r} \times \mathbf{p}$ is conserved so that $\mathbf{r} \cdot \mathbf{L} = 0$. Thus $\mathbf{r}$ always lies in a plane [4] whose normal is parallel to $\mathbf{L}$. Since the motion of the particle is planar because of symmetry considerations it is appropriate to ask: Suppose the motion of the particle was de facto planar and not due to symmetry considerations as in the case of motion under a central force, could one still obtain a Runge-Lenz vector that is conserved in this 2+1 dimensional case? Happily, we derive in this paper following Dahl's method [2] a bonafide Runge-Lenz vector (see eq. (23) below) associated with the relativistic Lagrangian with Chern-Simons interactions [5-7] for a two-particle system consisting of masses $m_1$ and $m_2$ with charges $e_1$ and $e_2$ respectively, given by

$$L = -m_1 c^2 \left(1 - \frac{|\dot{x}_1^2|}{c^2}\right)^{1/2} - m_2 c^2 \left(1 - \frac{|\dot{x}_2^2|}{c^2}\right)^{1/2} - \frac{e_1 e_2}{\pi c} \frac{\mathbf{k} \cdot \mathbf{r} \times \dot{\mathbf{r}}}{|\mathbf{r}|^2} \tag{3}$$

with $\mathbf{r} = \mathbf{x}_1 - \mathbf{x}_2$, $\mathbf{x}_1$ (and $\dot{x}_1$) and $\mathbf{x}_2$ (and $\dot{x}_2$) being the position (and velocity) vectors of the masses $m_1$ and $m_2$ and $\mathbf{k}$ a fictitious unit vector orthogonal to the plane of $\mathbf{x}_1$ and $\mathbf{x}_2$.

It is easy to obtain (3) starting with the electromagnetic potentials $A^\mu (\mathbf{x},t)$, $\mu = 0,1,2$ for Chern-Simons electrodynamics in the radiation gauge with $\nabla \cdot \mathbf{A} = 0$, namely,

$$A^0(x,t) = \frac{e}{2\pi c} \frac{(\mathbf{x} - \mathbf{R}(t)) \times \dot{\mathbf{R}}(t) \cdot \mathbf{k}}{|\mathbf{x} - \mathbf{R}(t)|^2} \tag{4a}$$

$$A(x,t) = \frac{e}{2\pi} \frac{(x - R(t)) \times k}{|x - R(t)|^2} \tag{4b}$$

the charge e being located at **R**(t). Needless to say so, eqs.(4) are the 2 + 1 dimensional counterparts of the corresponding 3 + 1 dimensional $A^\mu$ (**x**, t) given by eqs.(26.19) and (26.20) in Fock [8] for example.

Under an infinitesmal Lorentz transformation given by $\delta x = -vt + \frac{v \cdot x}{c^2} \dot{x}$, it is simple to check that the change in the Lagrangian L which is defined by

$$\delta L = \frac{\partial L}{\partial x_1} \cdot \delta x_1 + \frac{\partial L}{\partial x_2} \cdot \delta x_2 + \frac{\partial L}{\partial \dot{x}_1} \cdot \delta \dot{x}_1 + \frac{\partial L}{\partial \dot{x}_2} \cdot \delta \dot{x}_2$$

$$= \frac{d}{dt}\left( -m_1 x_1 \cdot v \gamma_1 - m_2 x_2 \cdot v \gamma_2 + \frac{e_1 e_2}{\pi c} \frac{k \cdot \delta r \times r}{|r|^2} \right) \tag{5}$$

without the use of the equations of motion for the two masses $m_1$ and $m_2$. Here $\gamma_1$ and $\gamma_2$ are given by $\left(1 - \frac{|\dot{x}_1|^2}{c^2}\right)^{1/2}$ and $\left(1 - \frac{|\dot{x}_2|^2}{c^2}\right)^{1/2}$ respectively. Since $\delta L$ works out to a total differential in (5) without the use of the equations of motion, it is clear that the action $S = \int dt\, L$ is unaffected by the transformation from $x \rightarrow x + \delta x$ thus making the Lagrangian (3) Lorentz invariant.

The constant of motion is now obtained by using the equations of motion for $m_1$ and $m_2$ to rewrite the first equality in (5) as

$$\delta L = \frac{d}{dt}\left( \frac{\partial L}{\partial \dot{x}_1} \cdot \delta x_1 + \frac{\partial L}{\partial \dot{x}_2} \cdot \delta x_2 \right) \tag{6}$$

From the second equality in (5) and (6) one thus obtains

$$\frac{d}{dt}\left(\frac{\partial L}{\partial \dot{x}_1}.\delta x_1 + \frac{\partial L}{\partial \dot{x}_2}.\delta x_2 + m_1 x_1.v\gamma_1 + m_2 x_2.v\gamma_2 - \frac{e_1 e_2}{\pi c}\frac{k.\delta r \times r}{|r|^2}\right) = 0 \qquad (7)$$

With the canonical momenta defined by $p_i = \frac{\partial L}{\partial \dot{x}_i}$ it is easy to obtain from (3)

$p_1 = m_1 \dot{x}_1 \gamma_1^{-1} + g$ and $p_2 = m_2 \dot{x}_2 \gamma_2^{-1} - g$, where $\pi c |r|^2 g = e_1 e_2 (r \times k)$ and thus rework

(7) as $\frac{d}{dt}(K.v) = 0$ with the constant of motion $K$ given by

$$K = -t P + m_1 x_1 \gamma_1^{-1} + m_2 x_2 \gamma_2^{-1} \qquad (8)$$

the total momentum $P$ being defined as $P = p_1 + p_2$. Eq. (8) is the counterpart of eq. (20) in Ref. (2); also while the latter is derived from the Darwin Lagrangian [8] for the electromagnetic N-body problem and is correct in the $1/c^2$ approximation, eq. (8) above is exact. Note that $K$ depends explicitly on time except in the centre-of-momentum system which is the Lorentz frame where $P = 0$.

For the Lagrangian (3) the Hamiltonian H and the angular momentum vector $L$ are given by

$$H = m_1 c^2 \gamma_1^{-1} + m_2 c^2 \gamma_2^{-1}, \qquad L = x_1 \times p_1 + x_2 \times p_2 \qquad (9a)$$

and in the $P = 0$ frame they work out to

$$H = m_1 c^2 f_1 + m_2 c^2 f_2, \qquad L = r \times p = k(r_1 p_2 - r_2 p_1) = kL_z \qquad (9b)$$

with $p_1$ and $p_2$ being the x- and y-components of the relative momentum vector $p$

defined by $p = \frac{m_2 p_1 - m_1 p_2}{m_1 + m_2}$.

In deriving the Hamiltonian in (9b) we go through the following steps:

i) Since $m_1 \dot{\mathbf{x}}_1 \gamma_1^{-1} = \mathbf{p}_1 - \mathbf{g}$, therefore $\gamma_1^{-2} = 1 + \dfrac{1}{m_1^2 c^2} |\mathbf{p}_1 - \mathbf{g}|^2$. One can now express $\mathbf{p}_1$ and $\mathbf{p}_2$ in terms of $\mathbf{p}$ and $\mathbf{P}$ via the relations

$$\mathbf{p}_1 = \frac{m_1}{m_1 + m_2}\mathbf{P} + \mathbf{p}, \qquad \mathbf{p}_2 = \frac{m_2}{m_1 + m_2}\mathbf{P} - \mathbf{p} \qquad (10)$$

and find that in the centre - of - momentum system with $\mathbf{P} = 0$

$$\gamma_1^{-1} = \left(1 + \frac{|\mathbf{B}|^2}{m_1^2 c^2}\right)^{1/2} \qquad (11a)$$

with $\mathbf{B} = \mathbf{p} - \mathbf{g}$.

ii) The above calculation can be repeated with $m_2 \dot{\mathbf{x}}_2 \gamma_2^{-1} = \mathbf{p}_2 + \mathbf{g}$ to obtain in the case when $\mathbf{P} = 0$ the result

$$\gamma_2^{-1} = \left(1 + \frac{|\mathbf{B}|^2}{m_2^2 c^2}\right)^{1/2} \qquad (11b)$$

We shall re-label $\gamma_1^{-1}$ and $\gamma_2^{-1}$ in eqs.(11) as $f_1$ and $f_2$ respectively below. Using these definitions one can now rework $\mathbf{K}$ in eq. (8) in the centre-of-momentum system with $\mathbf{P} = 0$ as

$$\mathbf{K} = (m_1 f_1 + m_2 f_2)\mathbf{R} + \mu(f_1 - f_2)\mathbf{r} \qquad (12)$$

with the reduced mass $\mu$ and the centre - of - mass position vector $\mathbf{R}$ defined by

$$\mu = \frac{m_1 m_2}{m_1 + m_2} \text{ and } \mathbf{R} = \frac{m_1 \mathbf{x}_1 + m_2 \mathbf{x}_2}{m_1 + m_2}.$$

As emphasized by Dahl[2] the c. m. vector $\mathbf{R}$ should now be eliminated from eq. (12) to make $\mathbf{K}$ a proper dynamical function and as in Ref. 2 we shall do this with the help of the the equation of motion for $\mathbf{R}$ in the $\mathbf{P} = 0$ limit, namely,

$$\frac{d\mathbf{R}}{dt} = \left(\frac{\partial H}{\partial \mathbf{P}}\right)_{P=0} = \left(\frac{\mu}{m_2}\frac{\partial H}{\partial \mathbf{p}_1} + \frac{\mu}{m_1}\frac{\partial H}{\partial \mathbf{p}_2}\right)_{p_1=p, p_2=-p} \tag{13}$$

One should now recast the Hamiltonian H in (9a) in terms of the canonical momenta $\mathbf{p}_1$ and $\mathbf{p}_2$ and it is given by

$$H = m_1 c^2 g_1 + m_2 c^2 g_2, \quad g_1 = \left(1 + \frac{|S_1|^2}{m_1^2 c^2}\right)^{1/2}, g_2 = \left(1 + \frac{|S_2|^2}{m_2^2 c^2}\right)^{1/2} \tag{14}$$

with $\mathbf{S}_1 = \mathbf{p}_1 - \mathbf{g}$ and $\mathbf{S}_2 = \mathbf{p}_2 + \mathbf{g}$. On calculating the partial derivatives required in eq. (13) it is easy to get

$$(m_1 + m_2)\frac{d\mathbf{R}}{dt} = (\gamma_1 - \gamma_2)\mathbf{B} \tag{15}$$

From the Hamiltonian (9b), the equations of motion are:

$$\frac{\partial H}{\partial \mathbf{p}} = T\mathbf{B} = \dot{\mathbf{r}} \tag{16a}$$

with $T = \frac{\gamma_1}{m_1} + \frac{\gamma_2}{m_2}$, $\gamma_1$ and $\gamma_2$ being defined by eqs.(11). Since $\mathbf{B} = \mathbf{p} - \mathbf{g}$ it is clear that

$$\mathbf{k} \times \dot{\mathbf{r}} = T(\mathbf{k} \times \mathbf{p} + h\mathbf{r}) \tag{16b}$$

where $\pi c |\mathbf{r}|^2 h = -e_1 e_2$. With eq. (16b) the other equation of motion, namely,

$$\frac{\partial H}{\partial \mathbf{r}} = T h\left(\mathbf{k} \times \mathbf{p} - \mathbf{r}\left(h + 2\frac{\mathbf{r}.\mathbf{k} \times \mathbf{p}}{|\mathbf{r}|^2}\right)\right) = -\dot{\mathbf{p}} \tag{16c}$$

yields

$$\dot{\mathbf{p}} = h\left(\mathbf{k} \times \dot{\mathbf{r}} - 2\frac{\mathbf{r}}{|\mathbf{r}|^2}\mathbf{k} \times \dot{\mathbf{r}}.\mathbf{r}\right) \tag{17}$$

Eq. (17) easily leads to

$$\frac{d}{dt}(p + h\mathbf{r} \times \mathbf{k}) = \frac{d\mathbf{B}}{dt} = \mathbf{0} \tag{18}$$

Thus **B** is a constant in time and so are $f_1$ and $f_2$ in eqs.(11). Returning to eq. (15) we get

$$\frac{d}{dt}\left(\mathbf{R} + \frac{\gamma_2 - \gamma_1}{m_1 + m_2}\mathbf{B}\right) = \mathbf{0} \tag{19}$$

on using eq. (18). Thus the c. m. position vector **R**(t) is given by

$$(m_1 + m_2)\mathbf{R}(t) = \mathbf{B}(\gamma_1 - \gamma_2)t + \mathbf{R}_0 \tag{20}$$

where $\mathbf{R}_0$ is a constant. Eq. (20) is the counterpart of eq. (43) in Ref. 2; however we must emphasize here that eq. (20) above is exact in that it has been derived using the equations of motion given by (16a) and (16c) without recourse to approximation. In contrast, the solution for **R**(t) given by eq. (43) in Dahl's paper [2] has been obtained in the $1/c^2$ approximation to the equation of motion as given by eq. (39) therein. On using (20) in (12) one obtains the exact form of the vector **K** as

$$\mathbf{K} = \left(\frac{m_1 - m_2 + m_2 f_2 \gamma_1 - m_1 f_1 \gamma_2}{m_1 + m_2}\right) t \mathbf{B} + \mu(f_1 - f_2)\mathbf{r} + \mathbf{R}_o \frac{m_1 f_1 + m_2 f_2}{m_1 + m_2} \tag{21}$$

With the Hamiltonian H in (9b), eq.(21) can be reworked as

$$\mathbf{K} = \frac{H}{(m_1 + m_2)c^2}\mathbf{R}_0 + \frac{m_1 - m_2}{m_1 + m_2}\mathbf{M} \tag{22}$$

with

$$\mathbf{M} = Qt\mathbf{B} + m_1 m_2 \frac{f_1 - f_2}{m_1 - m_2}\mathbf{r} \tag{23}$$

and $Q = 1 + \dfrac{m_2 f_2 \gamma_1 - m_1 f_1 \gamma_2}{m_1 - m_2}$. For the Chern-Simons Lagrangian given by eq.(3) above eq.(22) is the counterpart of eq.(46) in Ref.2 ;eq.(23) therefore defines the Runge-Lenz vector **M** of the Lagrangian (3) in the centre-of-momentum system.

It is easy to see that the above derivation in (22) and (23) needs a second look for the equal mass $m_1 = m_2$ case. Indeed a similar discussion is also in order for the Runge-Lenz vector derived from eq.(45) in Dahl's paper [2] ; note however that the remedy there is quite painless, namely: begin with unequal masses $m_1 \neq m_2$ and subsequently derive the correct Lenz vector for the equal mass case, $m_1 = m_2 = m$, the result being a simple replacement of the factor μ in eq.(46) there by m/2. But a corresponding effort here merits a separate discussion and is therefore relegated to the Appendix at the end of this paper.

With $M_1$ denoting the x-component of the 2-component vector **M** it is clear that $\dfrac{\partial M_1}{\partial t} = Q B_1$ ; thus one expects the Poisson bracket $\{M_1, H\} = -Q B_1$, since **M** is a constant of motion, with H being given by eq. (9b). It is easy to check this using the following:

$$\{r_1, f_1\} = \dfrac{\gamma_1 B_1}{m_1^2 c^2}, \{r_1, f_2\} = \dfrac{\gamma_2 B_1}{m_2^2 c^2}, \{r_i, p_j\} = \delta_{ij}, \{f_1, f_2\} = 0, \{\mathbf{B}, f_1\} = 0 = \{\mathbf{B}, f_2\} \quad (24$$

On replacing $r_1$ by $r_2$ in the first pair of Poisson brackets in (24) one should also replace $B_1$ by $B_2$.

One can also use eq. (24) to evaluate the Poisson bracket of **M** with $L_z$ the latter being given in eq. (9b). Using $\{B_i, L_z\} = -\varepsilon_{ij} B_j$ it is easy to verify that

$$\{M_i, L_z\} = -\varepsilon_{ij} M_j \qquad (25)$$

with $\varepsilon_{12} = -\varepsilon_{21} = 1$. We shall now evaluate the $\{M_1, M_2\}$ Poisson bracket; by virtue of eq. (24) and the fact that $\{B_1, B_2\} = 0$ it is easy to obtain the result

$$\{M_1, M_2\} = -\frac{\alpha^2}{c^2}(f_1 - f_2)\left(\frac{\gamma_1}{m_1^2} - \frac{\gamma_2}{m_2^2}\right)\left(\frac{e_1 e_2}{\pi c} + L_z\right) \qquad (26)$$

with $(m_1 - m_2)\alpha = m_1 m_2$. While eq. (25) is expected on account of **M** being a vector, (26) does not match with that given by eq. (2) for the Runge-Lenz vector **K** and is therefore a novel feature of this paper. The fact that the calculation made here is exact and quite unlike the the approximate calculation to order $1/c^2$ of the Runge-Lenz vector by Dahl [2] reinforces our confidence in this assertion.

The extra term $e_1 e_2/\pi c$ in eq. (26) is reminiscent of the observation by Witten and Olive [9,10] long ago that in supersymmetric theories with solitons the usual supersymmetry algebra is modified to include topological quantum numbers as central charges. Interestingly, the last factor on the right hand side of eq.(26) can be expressed as

$\frac{e_1 e_2}{\pi c} + L_z = \mu(\boldsymbol{r} \times \boldsymbol{l})_z$ where we define $\boldsymbol{l} = \dot{\boldsymbol{x}}_1 \gamma_1^{-1} - \dot{\boldsymbol{x}}_2 \gamma_2^{-1}$ with $\gamma_1 = \left(1 - \frac{|\dot{\boldsymbol{x}}_1|^2}{c^2}\right)^{1/2}$ and

$\gamma_2 = \left(1 - \frac{|\dot{\boldsymbol{x}}_2|^2}{c^2}\right)^{1/2}$. Qualitatively, eqs.(25) and (26) are important to this paper because we have derived them, besides eq.(24), as Poisson brackets here. This is not the case with the Poisson bracket relations given by eq.(4) in Ref.2; indeed Dahl [2] provides only a qualitative understanding (see eqs.(56) - (58) in Ref.(2)) of the said Poisson bracket and is unable to derive eq.(4) in the $1/c^2$ order description in his paper. Needless to say so,

eq.(26) is the 2 + 1 dimensional counterpart of the 3 + 1 dimensional Poisson bracket given by eq.(4) in Ref.(2), namely

$$\{M_i, M_j\} = -\frac{2}{\mu} H \sum_{k=1}^{3} \varepsilon_{ijk} L_k \qquad (27)$$

In conclusion, two distinguishing features characterise the derivation of the Runge-Lenz vector (23) associated with the Chern-Simons Lagrangian (3) in this paper: a) The calculation here is exact and, (b) as an unexpected bonus the Poisson bracket in (26) contains the product $e_1 e_2$ as a central charge.

Acknowledgements: I thank Suresh Govindarajan for reminding me of Ref. 9 and G. Date of the Institute of Mathematical Sciences, Chennai and P. K. Panigrahi of the School of Physics, University of Hyderabad for clarifying discussions. I also thank an anonymous referee for several suggestions to improve the presentation of the paper.

**Appendix**

When the masses $m_1$ and $m_2$ in the Lagrangian (3) are equal the Runge-Lenz vector **M** is worked out initially by assuming that $m_1 \neq m_2$; taking the limit of eq.(22) when $m_2 \to m_1$ then yields the desired form of **M** as explained below. Let us assume here that $m_1 = m_2 + \varepsilon$ where $\varepsilon > 0$. For the second term in eq.(23) of the text one thus has

$$\lim_{\varepsilon \to 0} \frac{1}{\varepsilon}(f_1 - f_2) = \lim_{\varepsilon \to 0} \left\{ \left(1 + \frac{|\mathbf{B}|^2}{c^2(m_2 + \varepsilon)^2}\right)^{1/2} - \left(1 + \frac{|\mathbf{B}|^2}{c^2 m_2^2}\right)^{1/2} \right\} \tag{A1}$$

$$= -\frac{|\mathbf{B}|^2}{c^2 m_2^3} \gamma_2 \tag{A2}$$

by a Maclaurin expansion of the first term to order $\varepsilon$. Here $\gamma_2$ and $\gamma_1$ (below) are given by eqs.(11) of the text. A similar effort on the first term in eq.(23) yields

$$\lim_{\varepsilon \to 0}\left(1 - \frac{m_1 f_1 \gamma_2 - m_2 f_2 \gamma_1}{\varepsilon}\right) = 2\gamma_2 \frac{|\mathbf{B}|^2}{m_2^2 c^2} \tag{A3}$$

Thus when $m_1 = m_2 = m$, we have instead of eq.(23) the Runge-Lenz vector

$$\mathbf{M} = \frac{|\mathbf{B}|^2 \gamma}{mc^2}\left(\frac{2t}{m}\mathbf{B} - \mathbf{r}\right) = \gamma(1 - f^2)(m\mathbf{r} - 2t\mathbf{B}) \tag{A4}$$

with $f\gamma = 1$ and $\gamma = \left(1 + \frac{|\mathbf{B}|^2}{m^2 c^2}\right)^{-1/2}$. With the Poisson brackets given by eqs.(24) in the text it is now easy to arrive at the relations $\{M_i, L_z\} = -\varepsilon_{ij} M_j$ and

$$\{M_1, M_2\} = -\frac{1}{c^2}\gamma^4(f^4 - 1)\left(\frac{e_1 e_2}{\pi c} + L_z\right) \tag{A5}$$

Eq.(A5) is the equal mass counterpart of eq.(26) of the text for unequal masses..